# Polarization of tightly focused laser beams

**John Lekner**

School of Chemical and Physical Sciences, Victoria University of Wellington, PO Box 600,
Wellington, New Zealand

**Abstract**
The polarization properties of monochromatic light beams are studied. In
contrast to the idealization of an electromagnetic plane wave, finite beams
which are everywhere linearly polarized in the same direction do not exist.
Neither do beams which are everywhere circularly polarized in a fixed plane.
It is also shown that transversely finite beams cannot be purely transverse in
both their electric and magnetic vectors, and that their electromagnetic
energy travels at less than $c$. The electric and magnetic fields in an
electromagnetic beam have different polarization properties in general, but
there exists a class of *steady beams* in which the electric and magnetic
polarizations are the same (and in which energy density and energy flux are
independent of time). Examples are given of exactly and approximately
linearly polarized beams, and of approximately circularly polarized beams.

(Some figures in this article are in colour only in the electronic version)

## 1. Introduction

An electromagnetic wave is specified in terms of the electric
and magnetic fields $\boldsymbol{E}$ and $\boldsymbol{B}$. For monochromatic waves of
angular frequency $\omega$ we can write [1, 2]

$$\boldsymbol{E}(\boldsymbol{r}, t) = \mathrm{Re}\{\boldsymbol{E}(\boldsymbol{r})\mathrm{e}^{-\mathrm{i}\omega t}\} = \boldsymbol{E}_r(\boldsymbol{r})\cos\omega t + \boldsymbol{E}_i(\boldsymbol{r})\sin\omega t$$
$$\boldsymbol{B}(\boldsymbol{r}, t) = \mathrm{Re}\{\boldsymbol{B}(\boldsymbol{r})\mathrm{e}^{-\mathrm{i}\omega t}\} = \boldsymbol{B}_r(\boldsymbol{r})\cos\omega t + \boldsymbol{B}_i(\boldsymbol{r})\sin\omega t \quad (1)$$

where $\boldsymbol{E}_r$ and $\boldsymbol{E}_i$ are the real and imaginary parts of the
complex vector $\boldsymbol{E}(\boldsymbol{r})$, and likewise for $\boldsymbol{B}(\boldsymbol{r})$. The polarization
properties of the wave usually refer to those of the electric
field. For the plane wave in vacuum, $\boldsymbol{E}(\boldsymbol{r}) = \boldsymbol{E}_0\mathrm{e}^{\mathrm{i}\boldsymbol{k}\cdot\boldsymbol{r}}$, we have
$\boldsymbol{B}(\boldsymbol{r}) = k^{-1}\boldsymbol{k} \times \boldsymbol{E}(\boldsymbol{r})$ where $k = \omega/c$ and $\boldsymbol{k}$ gives the direction
of propagation, so $\boldsymbol{E}$ and $\boldsymbol{B}$ have the same polarization
properties, but in general the polarization properties of $\boldsymbol{E}$ and
$\boldsymbol{B}$ will differ. There exists a special class of monochromatic
beams (*steady beams*, introduced in section 4 of [3] and studied
in more detail in section 4 of [4]) for which $\boldsymbol{E}(\boldsymbol{r}) = \pm\mathrm{i}\boldsymbol{B}(\boldsymbol{r})$
and for these beams the polarization properties of the electric
and magnetic fields are the same, as we shall see shortly.

At a fixed point in space, the endpoint of the vector $\boldsymbol{E}(\boldsymbol{r}, t)$
describes an ellipse in time $2\pi/\omega$ (see for example [2, section
1.4.3]): one can write

$$\boldsymbol{E}_r + \mathrm{i}\boldsymbol{E}_i = (\boldsymbol{E}_1 + \mathrm{i}\boldsymbol{E}_2)\mathrm{e}^{\mathrm{i}\gamma} \quad (2)$$

and $\gamma$ can be chosen so that the real vectors $\boldsymbol{E}_1$ and $\boldsymbol{E}_2$ are
perpendicular. This value of $\gamma$ is given by

$$\tan 2\gamma = \frac{2\boldsymbol{E}_r \cdot \boldsymbol{E}_i}{E_r^2 - E_i^2}. \quad (3)$$

Since $\boldsymbol{E}_1 = \boldsymbol{E}_r\cos\gamma + \boldsymbol{E}_i\sin\gamma$ and $\boldsymbol{E}_2 = -\boldsymbol{E}_r\sin\gamma +
\boldsymbol{E}_i\cos\gamma$, in the plane of $(\boldsymbol{E}_r, \boldsymbol{E}_i)$ or $(\boldsymbol{E}_1, \boldsymbol{E}_2)$ the electric field

$$\boldsymbol{E}(\boldsymbol{r}, t) = \mathrm{Re}\{(\boldsymbol{E}_1 + \mathrm{i}\boldsymbol{E}_2)\mathrm{e}^{\mathrm{i}(\gamma - \omega t)}\}$$
$$= \boldsymbol{E}_1\cos(\omega t - \gamma) + \boldsymbol{E}_2\sin(\omega t - \gamma) \quad (4)$$

has orthogonal components $\boldsymbol{E}_1$ and $\boldsymbol{E}_2$ (when $\gamma$ satisfies (3))
with magnitudes given by

$$\begin{pmatrix} E_1^2 \\ E_2^2 \end{pmatrix} = \tfrac{1}{2}\left[ E_r^2 + E_i^2 \pm \sqrt{(E_r^2 - E_i^2)^2 + 4(\boldsymbol{E}_r \cdot \boldsymbol{E}_i)^2} \right]. \quad (5)$$

$E_1$ and $E_2$ give the lengths of the semiaxes of the polarization
ellipse. $E_2$ is zero for *linear polarization*, for which we
therefore need $\boldsymbol{E}_r$ and $\boldsymbol{E}_i$ to be collinear:

$$E_r^2 E_i^2 - (\boldsymbol{E}_r \cdot \boldsymbol{E}_i)^2 = 0 \qquad \text{(linear polarization)}. \quad (6)$$

The condition for $\boldsymbol{E}_r$ and $\boldsymbol{E}_i$ to be collinear can also be written
as $\boldsymbol{E}_r \times \boldsymbol{E}_i = 0$. The square of this relation reduces to (6).

$E_1^2 = E_2^2$ for *circular polarization*, for which we need $\boldsymbol{E}_r$
and $\boldsymbol{E}_i$ to be perpendicular and equal in magnitude:

$$\{\boldsymbol{E}_r \cdot \boldsymbol{E}_i = 0 \text{ and } E_r^2 = E_i^2\} \qquad \text{(circular polarization)}. \quad (7)$$

We can define a degree of linear polarization $\Lambda(r)$, depending on position within the monochromatic coherent beam under consideration; for the electric polarization this is

$$\Lambda = \frac{[(E_r^2 - E_i^2)^2 + 4(\boldsymbol{E}_r \cdot \boldsymbol{E}_i)^2]^{\frac{1}{2}}}{E_r^2 + E_i^2} = \frac{|\boldsymbol{E}^2(r)|}{|\boldsymbol{E}(r)|^2} \quad (8)$$

with an analogous expression for the magnetic polarization. When the real and imaginary parts of $\boldsymbol{E}(r) = \boldsymbol{E}_r(r) + \mathrm{i}\boldsymbol{E}_i(r)$ are collinear, which is the condition (6) for linear polarization, $\Lambda$ is unity. When $\boldsymbol{E}_r$ and $\boldsymbol{E}_i$ are orthogonal and equal in magnitude, the conditions for circular polarization (7), $\Lambda$ is zero. Note that the conditions (7) for circular polarization can be condensed into $\boldsymbol{E}^2 = 0$ for the complex field $\boldsymbol{E} = \boldsymbol{E}_r + \mathrm{i}\boldsymbol{E}_i$: the complex field is nilpotent on the curve $C$ where it is circularly polarized. (A measure of the degree of circular polarization is $1 - \Lambda$.) From (5) we have

$$\Lambda = \frac{E_1^2 - E_2^2}{E_1^2 + E_2^2} = \frac{e^2}{2 - e^2} \quad (9)$$

where $e$ is the eccentricity of the polarization ellipse, given by $e^2 = 1 - (E_2/E_1)^2$. Thus $e^2$ has the same limiting values of unity and zero for linear and circular polarization as does $\Lambda$: $e^2 = 2\Lambda/(1 + \Lambda)$.

For steady beams, within which both the energy flux (Poynting vector) $\boldsymbol{S}(r, t) = \frac{c}{4\pi}\boldsymbol{E}(r, t) \times \boldsymbol{B}(r, t)$ and the energy density $u(r, t) = \frac{1}{8\pi}(E^2(r, t) + B^2(r, t))$ are independent of time, we have [3, 4] $\boldsymbol{E}(r) = \pm \mathrm{i}\boldsymbol{B}(r)$ so that $\boldsymbol{B}_r = \pm \boldsymbol{E}_i$ and $\boldsymbol{B}_i = \mp \boldsymbol{E}_r$. For electromagnetic steady beams, therefore, the electric and magnetic polarization properties are identical, since the angle $\gamma$ is the same for $\boldsymbol{E}(r)$ and $\boldsymbol{B}(r)$, and $E_1^2 = B_1^2$, $E_2^2 = B_2^2$.

A simple example of differing electric and magnetic polarization properties is provided by electric dipole radiation. The complex fields are (see e.g. [1], section 9.2)

$$\boldsymbol{B}(r) = k^2(\hat{r} \times \boldsymbol{p})\left(1 + \frac{\mathrm{i}}{kr}\right)\frac{\mathrm{e}^{\mathrm{i}kr}}{r}$$

$$\boldsymbol{E}(r) = k^2(\hat{r} \times \boldsymbol{p}) \times \hat{r}\frac{\mathrm{e}^{\mathrm{i}kr}}{r} + [3(\hat{r} \cdot \boldsymbol{p})\hat{r} - \boldsymbol{p}]\left(\frac{1}{r^3} - \frac{\mathrm{i}k}{r^2}\right)\mathrm{e}^{\mathrm{i}kr} \quad (10)$$

where $\boldsymbol{p}$ is the electric dipole moment, and $\hat{r} = \boldsymbol{r}/r$. Thus the real and imaginary parts of $\boldsymbol{B}(r)$ are collinear,

$$\boldsymbol{B}_r(r) = k^2(\hat{r} \times \boldsymbol{p})\frac{1}{r}\left[\cos kr - \frac{\sin kr}{kr}\right]$$
$$\boldsymbol{B}_i(r) = k^2(\hat{r} \times \boldsymbol{p})\frac{1}{r}\left[\sin kr + \frac{\cos kr}{kr}\right] \quad (11)$$

and therefore the magnetic field is everywhere linearly polarized (along the direction of $\hat{r} \times \boldsymbol{p}$), but the electric field is in general elliptically polarized, with only one surface on which the polarization is exactly linear, as we shall see below.

Nisbet and Wolf [5] have considered electromagnetic waves within which one of the field vectors is everywhere linearly polarized. The direction of polarization is fixed at a given point in space, but 'may be different at different points in the field'. In fact we shall show in the next section that a finite beam with either $\boldsymbol{E}$ or $\boldsymbol{B}$ everywhere linearly polarized in the same fixed direction cannot exist.

As an example of polarization in monochromatic light beams, consider the TM (transverse magnetic) beams [3, 6] characterized by a vector potential aligned with the beam axis,

$$\boldsymbol{A}(r) = A_0(0, 0, \psi) \quad (12)$$

where $\psi$ is a solution of the Helmholtz equation $(\nabla^2 + k^2)\psi = 0$. These beams have $\boldsymbol{B}(r) = \nabla \times \boldsymbol{A}(r) = A_0\left(\frac{\partial \psi}{\partial y}, -\frac{\partial \psi}{\partial x}, 0\right)$, transverse to the propagation ($z$) direction. When $\psi$ is independent of the azimuthal angle $\phi$, the complex fields are

$$\boldsymbol{B}(r) = A_0\left(\frac{\partial \psi}{\partial \rho}\sin\phi, -\frac{\partial \psi}{\partial \rho}\cos\phi, 0\right)$$
$$\boldsymbol{E}(r) = \frac{\mathrm{i}A_0}{k}\left(\frac{\partial^2 \psi}{\partial \rho \partial z}\cos\phi, \frac{\partial^2 \psi}{\partial \rho \partial z}\sin\phi, \frac{\partial^2 \psi}{\partial z^2} + k^2\psi\right) \quad (13)$$

where $\rho = (x^2 + y^2)^{\frac{1}{2}}$ is the distance from the beam axis. If we write the complex wavefunction $\psi(\rho, z)$ as $\psi_r + \mathrm{i}\psi_i$, the real and imaginary parts of $\boldsymbol{B}(r)$ are (we take $A_0$ to be real)

$$\boldsymbol{B}_{r,i}(r) = A_0\frac{\partial \psi_{r,i}}{\partial \rho}(\sin\phi, -\cos\phi, 0) \quad (14)$$

and thus $\boldsymbol{B}_r$ and $\boldsymbol{B}_i$ are collinear, and the magnetic field is everywhere linearly polarized. (The magnetic field lines are circles, with centres on the beam axis.) The electric field, on the other hand, has

$$\boldsymbol{E}_{r,i}(r) = \mp\frac{A_0}{k}\left(\frac{\partial^2 \psi_{i,r}}{\partial \rho \partial z}\cos\phi, \frac{\partial^2 \psi_{i,r}}{\partial \rho \partial z}\sin\phi, \frac{\partial^2 \psi_{i,r}}{\partial z^2} + k^2\psi_{i,r}\right) \quad (15)$$

and is thus elliptically polarized, in general. Note that $\boldsymbol{E}$ has a longitudinal component, the necessity of which was noted in [7]. In fact we shall see in the next section that finite beams cannot be purely transverse in both electric and magnetic fields.

Nye and Hajnal [8–12] have studied and classified the polarization of electromagnetic waves by means of their geometric properties. The location of circular polarization of the electric field of a monochromatic wave is specified by (7). These two conditions, $\boldsymbol{E}_r \cdot \boldsymbol{E}_i = 0$ and $E_r^2 = E_i^2$, each determine a surface in space, and the two surfaces intersect on a curve $C$ on which the electric field is circularly polarized. Likewise, the location of linear polarization is determined by one condition (6), namely $(\boldsymbol{E}_r \cdot \boldsymbol{E}_i)^2 = E_r^2 E_i^2$. This equation, or equivalently the condition that $\boldsymbol{E}_r$ and $\boldsymbol{E}_i$ be collinear (which includes as a special case that one of them be zero), determines a surface $S$ in space. In a given plane, e.g. $z =$ constant, the $S$ surface intersects a curve on which the polarization is linear, although the field direction in general varies along the curve. The curve $C$ does not cross an $S$ surface (the field cannot be both circularly and linearly polarized at any point), except possibly at a point where the field is zero.

By way of example, the electric field of a radiating electric dipole, given in (10), is never circularly polarized. The condition for linear polarization is that $E_r^2 E_i^2 - (\boldsymbol{E}_r \cdot \boldsymbol{E}_i)^2$ be zero. This expression factors to $4p^4 k^6 r^{-6} \cos^2\theta \sin^2\theta$, where $\theta$ is the angle between $\boldsymbol{p}$ and $\boldsymbol{r}$. Thus the electric field is linearly polarized in the equatorial plane $\theta = \pi/2$, where it takes the value $\boldsymbol{E} = pr^{-3}[(kr)^2 + \mathrm{i}kr - 1]\mathrm{e}^{\mathrm{i}kr}$. The electric field is also linearly polarized on the axis of the dipole, where $\boldsymbol{E} = 2pr^{-3}(1 - \mathrm{i}kr)\mathrm{e}^{\mathrm{i}kr}$. The equatorial plane and the polar

J Lekner

axis may be viewed as parts of the same surface, obtained by rotating the 'four-leaf clover' $r^6 = a^6 \cos^2\theta \sin^2\theta$ about the polar axis $\theta = 0$, in the limit as $a \to \infty$.

In sections 3 and 4 we shall examine the polarization properties of some special beam wavefunctions. Before that, section 2 gives general results relating to beams.

## 2. Non-existence theorems for electromagnetic beams

The textbook electromagnetic plane wave, in which $\boldsymbol{E}$, $\boldsymbol{B}$ and the propagation vector $\boldsymbol{k}$ are everywhere mutually perpendicular, can be everywhere linearly polarized in the same direction, or everywhere circularly polarized in the same plane, and its energy is everywhere transported in a fixed direction at the speed of light. We shall show that none of these properties can hold for transversely finite monochromatic electromagnetic beams.

### 2.1. Pure TEM beam modes do not exist

Suppose $\boldsymbol{E} = (E_x, E_y, 0)$, $\boldsymbol{B} = (B_x, B_y, 0)$. With time-dependence $e^{-ickt}$, the Maxwell curl equations in free space become $\nabla \times \boldsymbol{E} = ik\boldsymbol{B}$, $\nabla \times \boldsymbol{B} = -ik\boldsymbol{E}$. The first gives $ik(B_x, B_y, 0) = \left(-\frac{\partial E_y}{\partial z}, \frac{\partial E_x}{\partial z}, \frac{\partial E_y}{\partial x} - \frac{\partial E_x}{\partial y}\right)$. Taking the curl gives $ik\nabla \times \boldsymbol{B} = \left(-\frac{\partial^2 E_x}{\partial z^2}, -\frac{\partial^2 E_y}{\partial z^2}, \frac{\partial}{\partial z}\left(\frac{\partial E_x}{\partial x} + \frac{\partial E_y}{\partial y}\right)\right) = k^2(E_x, E_y, 0)$. Thus we have $\frac{\partial E_y}{\partial x} = \frac{\partial E_x}{\partial y}$, $\frac{\partial E_x}{\partial x} + \frac{\partial E_y}{\partial y} = 0$ (from $\nabla \cdot \boldsymbol{E} = 0$) and $\frac{\partial^2 E_x}{\partial z^2} + k^2 E_x = 0$, $\frac{\partial^2 E_y}{\partial z^2} + k^2 E_y = 0$. Propagating solutions of the last two equations are of the form $E_x = e^{ikz} F(x, y)$, $E_y = e^{ikz} G(x, y)$, and the preceding two equations then become $\frac{\partial G}{\partial x} = \frac{\partial F}{\partial y}$, $\frac{\partial F}{\partial x} + \frac{\partial G}{\partial y} = 0$. Thus $F$ and $G$ are harmonic functions, $\left(\frac{\partial^2}{\partial x^2} + \frac{\partial^2}{\partial y^2}\right) F, G = 0$; hence $F$ and $G$ cannot have maxima or minima, and so cannot localize the field around the beam axis. (TEM modes in waveguides can exist, in the presence of two or more cylindrical conductors: see for example section 8.2 of [1].)

### 2.2. Beams of fixed linear polarization do not exist

Suppose $\boldsymbol{E} = (F, 0, 0)$. Then from $\nabla \times \boldsymbol{E} = ik\boldsymbol{B}$, $\nabla \times \boldsymbol{B} = -ik\boldsymbol{E}$ we have $ik\boldsymbol{B} = \left(0, \frac{\partial F}{\partial z}, -\frac{\partial F}{\partial y}\right)$, $ik\nabla \times \boldsymbol{B} = \nabla \times (\nabla \times \boldsymbol{E}) = k^2\boldsymbol{E}$, which gives $\frac{\partial^2 F}{\partial y^2} + \frac{\partial^2 F}{\partial z^2} + k^2 F = 0$, $\frac{\partial^2 F}{\partial x \partial y} = 0 = \frac{\partial^2 F}{\partial x \partial z}$. The last two equations imply that $F = F(y, z)$, and thus a beam everywhere polarized along the $x$-direction cannot be localized in the $x$-direction.

### 2.3. Beams which are everywhere circularly polarized in a fixed plane do not exist

The conditions for circular polarization, equations (7), state that $\boldsymbol{E}_r$ and $\boldsymbol{E}_i$ be everywhere perpendicular, and of equal magnitude. If $\boldsymbol{E}$ lies in the $xy$ plane, we can take $\boldsymbol{E}_r = (F(\boldsymbol{r}), 0, 0)$, $\boldsymbol{E}_i = (0, F(\boldsymbol{r}), 0)$. For monochromatic waves, the curl equations of Maxwell give us, with $\omega = ck$,

$$\nabla \times \boldsymbol{E}_r + k\boldsymbol{B}_i = 0, \qquad \nabla \times \boldsymbol{E}_i - k\boldsymbol{B}_r = 0 \qquad (16)$$

$$\nabla \times \boldsymbol{B}_r - k\boldsymbol{E}_i = 0, \qquad \nabla \times \boldsymbol{B}_i + k\boldsymbol{E}_r = 0. \qquad (17)$$

The equations (16) give us $\boldsymbol{B}_i = k^{-1}\left(0, -\frac{\partial F}{\partial z}, \frac{\partial F}{\partial y}\right)$, $\boldsymbol{B}_r = k^{-1}\left(-\frac{\partial F}{\partial z}, 0, \frac{\partial F}{\partial x}\right)$, and then (17) give, respectively,

$$\frac{\partial^2 F}{\partial x^2} + \frac{\partial^2 F}{\partial z^2} + k^2 F = 0, \qquad \frac{\partial^2 F}{\partial x \partial y} = 0 = \frac{\partial^2 F}{\partial y \partial z}$$
$$\text{so } F = F(x, z) \qquad (18)$$

$$\frac{\partial^2 F}{\partial y^2} + \frac{\partial^2 F}{\partial z^2} + k^2 F = 0, \qquad \frac{\partial^2 F}{\partial x \partial y} = 0 = \frac{\partial^2 F}{\partial x \partial z},$$
$$\text{so } F = F(y, z). \qquad (19)$$

Thus $F$ must be a function of $z$ only, i.e. only plane waves can be everywhere circularly polarized in a fixed plane.

### 2.4. Beams or pulses within which the energy velocity is everywhere in the same direction and of magnitude $c$ do not exist

Suppose the energy velocity $\boldsymbol{v}_e = 2c\boldsymbol{E} \times \boldsymbol{B}/(E^2 + B^2)$ has magnitude $c$ everywhere. Then from equation (10) of [4] it follows that $E^2 = B^2$ and $\boldsymbol{E} \cdot \boldsymbol{B} = 0$, everywhere. Let $\boldsymbol{E} \times \boldsymbol{B}$ point in the $z$ direction. Then $E_y B_z = B_y E_z$, $E_z B_x = B_z E_x$ and $E_x B_y - B_x E_y = (E^2 + B^2)/2$. These three conditions, together with $E^2 = B^2$ and $\boldsymbol{E} \cdot \boldsymbol{B} = 0$ have only one real solution set, namely $\{E_x = B_y, E_y = -B_x, E_z = 0 = B_z\}$. This solution set is consistent with Maxwell's equations, provided that

$$\frac{\partial B_x}{\partial z} + \frac{1}{c}\frac{\partial B_x}{\partial t} = 0, \qquad \frac{\partial B_y}{\partial z} + \frac{1}{c}\frac{\partial B_y}{\partial t} = 0 \qquad (20)$$

$$\frac{\partial B_y}{\partial x} - \frac{\partial B_x}{\partial y} = 0, \qquad \frac{\partial B_x}{\partial x} + \frac{\partial B_y}{\partial y} = 0. \qquad (21)$$

Equations (20) are satisfied by $B_x$ and $B_y$ being arbitrary functions of $z - ct$. Equations (21) imply that $B_x$ and $B_y$ are harmonic functions in $x$ and $y$, so that the beam cannot be localized in the $x$ or $y$ directions. The only beams or pulses in which energy is everywhere transported in the same direction at the speed $c$ are unbounded in the transverse directions.

## 3. Examples of exactly and approximately linearly polarized beams

We have already seen, in section 1, that the TM beam with vector potential $\boldsymbol{A} = A_0(0, 0, \psi)$ is exactly linearly polarized in $\boldsymbol{B}$. Its dual is the TE beam, with $\boldsymbol{A} = (ik)^{-1} A_0\left(\frac{\partial \psi}{\partial y}, -\frac{\partial \psi}{\partial x}, 0\right)$ (see [3], section 4) for which the electric field is $\boldsymbol{E} = A_0\left(\frac{\partial \psi}{\partial y}, -\frac{\partial \psi}{\partial x}, 0\right)$ and thus an axially symmetric $\psi$ gives a beam within which the electric field is exactly linearly polarized everywhere. The electric field lines are circles concentric with the beam axis; the direction of polarization varies with the azimuthal angle. Note that setting $\psi = e^{ikz}$ gives identically zero fields: there is no plane wave limit for this class of beams. Figure 1 shows the electric field lines for this beam.

Consider now the fields in an approximately linearly polarized 'LP' beam resulting from the vector potential $\boldsymbol{A} = A_0(\psi, 0, 0)$: these are

$$\boldsymbol{B} = \nabla \times \boldsymbol{A} = A_0\left(0, \frac{\partial \psi}{\partial z}, -\frac{\partial \psi}{\partial y}\right) \qquad (22)$$

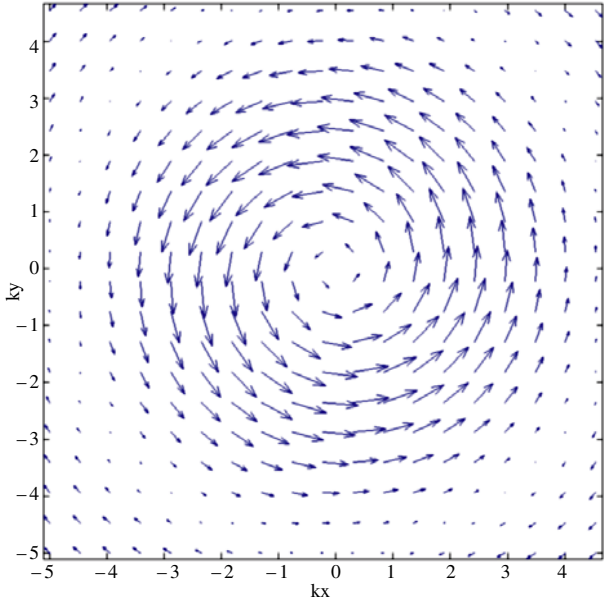

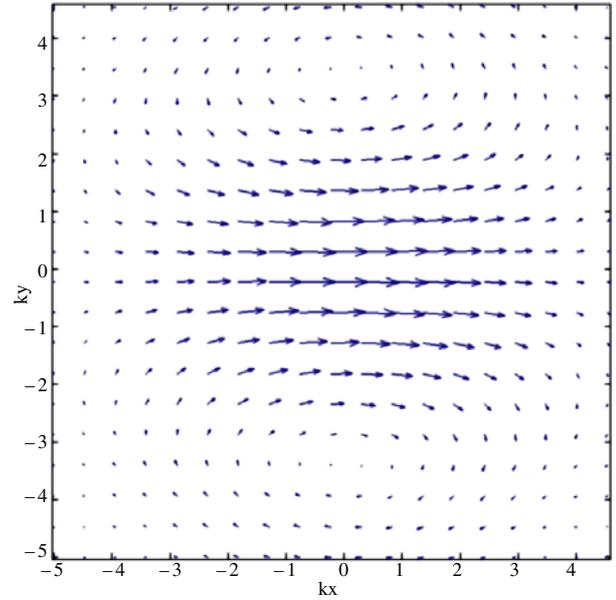

**Figure 1.** Electric field lines at $t = 0$ in the focal plane of the TE beam, plotted for the $\psi_{00} = \sin kR/kR$ wavefunction. The parameter $\beta = kb$ has the value 2, which corresponds to a tightly focused beam, with a beam divergence half-angle of $45°$. The electric field is everywhere purely transverse, and everywhere linearly polarized. The electric field is zero when $\partial \psi/\partial \rho$ is zero. For $\psi_{00}$ this is when $\tan kR = kR$. The first zero occurs at $kR \approx 4.49$; when $kb = 2$ this gives $k\rho \approx 5.48$.

**Figure 2.** Electric field at $\omega t = \pi/2$ for the 'LP' beam with vector potential directed along the $x$ axis, $\mathbf{A} = A_0(\psi, 0, 0)$, drawn for $\beta = 2$ in the focal plane. At $t = 0$ the field in the focal plane is purely longitudinal, at $\omega t = \pi/2$ it is purely transverse. The electric field is exactly linearly polarized in the $x = 0$ plane, where it is also exactly transverse, along $\hat{x}$. The transverse field has zeros on the $x$-axis, when $\frac{\partial^2 \psi}{\partial \rho^2} + k^2 \psi = 0$ (when $\beta = 2$ and $\psi = \psi_{00}$ the first zeros are at $k|x| \approx 5.23$); there the electric field is purely longitudinal, and is again linearly polarized. The field is null at $k|y| \approx 3.4$: see the caption to figure 3.

and
$$\mathbf{E} = \frac{\mathrm{i}}{k}\nabla(\nabla \cdot \mathbf{A}) + \mathrm{i}k\mathbf{A} = \frac{\mathrm{i}A_0}{k}\left(\frac{\partial^2 \psi}{\partial x^2} + k^2\psi, \frac{\partial^2 \psi}{\partial x \partial y}, \frac{\partial^2 \psi}{\partial x \partial z}\right). \tag{23}$$

In the plane wave limit $\psi \to \mathrm{e}^{\mathrm{i}kz}$ we have $\mathbf{B} \to \mathrm{i}kA_0\mathrm{e}^{\mathrm{i}kz}(0, 1, 0)$, $\mathbf{E} \to \mathrm{i}kA_0\mathrm{e}^{\mathrm{i}kz}(1, 0, 0)$; this has $\mathbf{E}$ and $\mathbf{B}$ both linearly polarized along mutually perpendicular transverse directions. We shall consider finite beams with $\psi$ independent of the azimuthal angle $\phi$. We then have $\psi = \psi(\rho, z)$ and the complex fields are

$$\mathbf{B} = A_0\left(0, \frac{\partial \psi}{\partial z}, -\sin\phi \frac{\partial \psi}{\partial \rho}\right)$$
$$\mathbf{E} = \frac{\mathrm{i}A_0}{k}\left(\cos^2\phi \frac{\partial^2 \psi}{\partial \rho^2} + \frac{\sin^2\phi}{\rho}\frac{\partial \psi}{\partial \rho} + k^2\psi, \right. \tag{24}$$
$$\left. \cos\phi\sin\phi\left(\frac{\partial^2 \psi}{\partial \rho^2} - \frac{1}{\rho}\frac{\partial \psi}{\partial \rho}\right), \cos\phi\frac{\partial^2 \psi}{\partial \rho \partial z}\right).$$

The beam wavefunction is complex: examples are the approximate Gaussian fundamental [13–15]

$$\psi_G = \frac{b}{b+\mathrm{i}z}\exp\left[\mathrm{i}kz - \frac{k\rho^2}{2(b+\mathrm{i}z)}\right] \tag{25}$$

and a set of exact complex source/sink solutions [3, 16, 17]

$$\psi_{\ell m} = j_\ell(kR)P_{\ell m}\left(\frac{z-\mathrm{i}b}{R}\right)\mathrm{e}^{\pm\mathrm{i}m\phi}$$
$$R = (z - \mathrm{i}b)\left[1 + \rho^2/(z - \mathrm{i}b)^2\right]^{\frac{1}{2}}. \tag{26}$$

Thus the fields given in (24) are clearly not linearly polarized everywhere, except in the plane wave limit. $\mathbf{B}$ is linearly polarized along $\hat{y}$ in the $y = 0$ plane ($\sin\phi = 0$), while $\mathbf{E}$ is linearly polarized along $\hat{x}$ in the $x = 0$ plane ($\cos\phi = 0$).

Figure 2 shows the electric field lines for the 'LP' beam with vector potential $\mathbf{A} = A_0(\psi, 0, 0)$, and figure 3 gives the contours of constant $\Lambda$. We see from (24) that the electric field is exactly linearly polarized (along $\hat{x}$) in the $x = 0$ plane. When $\psi$ is independent of $\phi$, the polarization measure for the electric field is

$$\Lambda = \left(\left|\left(\frac{1}{\rho}\psi_\rho + k^2\psi\right)^2 + \cos^2\phi\left\{\psi_{\rho\rho}(\psi_{\rho\rho} + 2k^2\psi)\right.\right.\right.$$
$$\left.\left.\left. - \frac{1}{\rho}\psi_\rho\left(\frac{1}{\rho}\psi_\rho + 2k^2\psi\right) + \psi_{\rho z}^2\right\}\right|\right)$$
$$\times \left(\left|\frac{1}{\rho}\psi_\rho + k^2\psi\right|^2 + \cos^2\phi\left\{|\psi_{\rho\rho}|^2 + |\psi_{\rho z}|^2 - \frac{1}{\rho^2}|\psi_\rho|^2\right.\right.$$
$$\left.\left. + 2k^2\operatorname{Re}\left[\left(\psi_{\rho\rho}^* - \frac{1}{\rho}\psi_\rho^*\right)\psi\right]\right\}\right)^{-1} \tag{27}$$

where subscripts denote differentiations with respect to $\rho$ or $z$. Appendix A gives further results when $\psi = \psi_{00}$.

## 4. Examples of beams which are approximately circularly polarized

We wish to construct beams which in the plane wave limit have the complex fields

$$\mathbf{E}(\mathbf{r}) = E_0\mathrm{e}^{\mathrm{i}kz}(\hat{x} \pm \mathrm{i}\hat{y}), \qquad \mathbf{B} = \mp\mathrm{i}\mathbf{E} \tag{28}$$

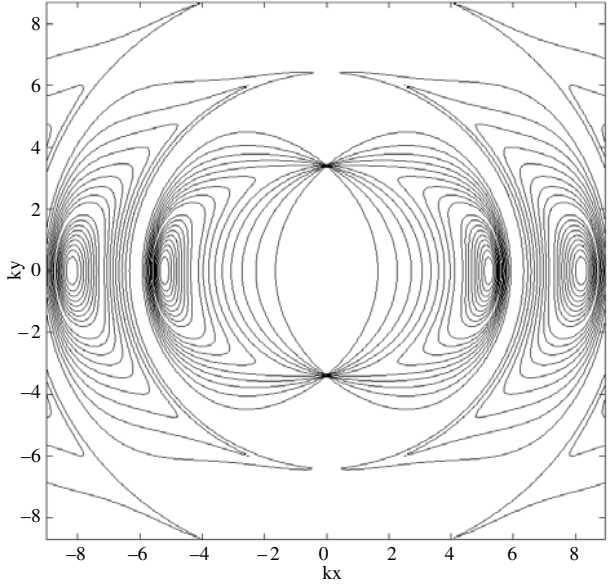

**Figure 3.** Contour plot of the polarization measure $\Lambda = |\boldsymbol{E}^2|/|\boldsymbol{E}|^2$ for the 'LP' beam with $\boldsymbol{A} = A_0(\psi, 0, 0)$, drawn for $\psi = \psi_{00}$ and $\beta = 2$ in the focal plane $z = 0$. The central region has $\Lambda > 0.9$, i.e. nearly linear polarization. The contours are at 10% intervals, from $\Lambda = 0.9$ to $0.1$. Going from the centre of the focal plane (where $\Lambda = 1$) out along the $x$-axis, the $\Lambda$ values first decrease steadily to zero (the first curves of circular polarization intersect the $x$-axis at $k|x| \approx 4.25$) and $\Lambda$ then increases again to unity (perfect linear polarization) at $k|x| \approx 5.23$. The cycle then repeats, with a decrease till the other side of the circular polarization curve is met at $k|x| \approx 5.62$. The next perfect linear polarization is at $k|x| \approx 8.02$ (all numbers are for $\psi_{00}$ with $\beta = 2$). Along the $y$-axis the polarization is linear everywhere, since there the electric field is proportional to $\left(\rho^{-1}\frac{\partial \psi}{\partial \rho} + k^2\psi, 0, 0\right)$. Since the field can be zero, for $\psi_{00}$ at points such that $\cot X = X^{-1} - X$ where $X = \beta\xi = \sqrt{(k\rho)^2 - \beta^2}$, other contours can touch at these null points, the first being at $k|y| \approx 3.4$.

(the upper sign corresponds to positive helicity) or the equivalent real fields

$$\boldsymbol{E}(\boldsymbol{r}, t) = E_0(\cos(kz - \omega t), \mp \sin(kz - \omega t), 0)$$
$$\boldsymbol{B}(\boldsymbol{r}, t) = E_0(\pm \sin(kz - \omega t), \cos(kz - \omega t), 0) \quad (29)$$

with the corresponding parts $\boldsymbol{E}_r = E_0(\cos kz, \mp \sin kz, 0)$, $\boldsymbol{E}_i = E_0(\sin kz, \pm \cos kz, 0)$; $\boldsymbol{B}_r = E_0(\pm \sin kz, \cos kz, 0)$, $\boldsymbol{B}_i = E_0(\mp \cos kz, \sin kz, 0)$.

Consider first the vector potential $\boldsymbol{A}_1 = k^{-1}E_0(-\mathrm{i}\psi, \psi, 0)$, where $\psi$ is some solution of the Helmholtz equation, for example one of the set (26). The plane wave limit, $\psi \to \mathrm{e}^{\mathrm{i}kz}$, then reproduces the positive helicity (upper sign) of (28). For general $\psi$ the fields are

$$\boldsymbol{B} = k^{-1}E_0\left(-\frac{\partial\psi}{\partial z}, -\mathrm{i}\frac{\partial\psi}{\partial z}, \chi\right), \quad \chi = \left(\frac{\partial}{\partial x} + \mathrm{i}\frac{\partial}{\partial y}\right)\psi \quad (30)$$

$$\boldsymbol{E} = E_0\left(\psi + k^{-2}\frac{\partial\chi}{\partial x}, \mathrm{i}\psi + k^{-2}\frac{\partial\chi}{\partial y}, k^{-2}\frac{\partial\chi}{\partial z}\right). \quad (31)$$

We see that for a finite beam we have lost full circular polarization, and also the *steady beam* property $\boldsymbol{E} = \pm\mathrm{i}\boldsymbol{B}$ which guarantees both the time invariance of energy flux and energy density [3, 4], and identical polarization properties for the electric and magnetic fields (section 1).

To obtain a steady approximately circularly polarized 'CP' beam we use the vector potential

$$\boldsymbol{A} = \tfrac{1}{2}[\boldsymbol{A}_1 + k^{-1}\nabla \times \boldsymbol{A}_1] = \tfrac{1}{2}k^{-2}E_0(-\Phi, -\mathrm{i}\Phi, \chi)$$
$$\Phi = \left(\frac{\partial}{\partial z} + \mathrm{i}k\right)\psi. \quad (32)$$

This gives the complex magnetic field

$$\boldsymbol{B} = \nabla \times \boldsymbol{A} = \left(\frac{E_0}{2k^2}\right)(\chi_y + \mathrm{i}\Phi_z, -\chi_x - \Phi_z, -\mathrm{i}(\Phi_x + \mathrm{i}\Phi_y)) \quad (33)$$

where the subscripts denote differentiations. The electric field is $\boldsymbol{E} = \mathrm{i}\boldsymbol{B}$ by construction. The plane wave limit again gives the circularly polarized fields of (28). We shall consider fields in which $\psi$ is independent of the azimuthal angle $\phi$, $\psi = \psi(\rho, z)$. Then, converting to cylindrical coordinates,

$$\chi = \mathrm{e}^{\mathrm{i}\phi}\psi_\rho$$
$$\chi_x = \mathrm{e}^{\mathrm{i}\phi}\left(\cos\phi\,\psi_{\rho\rho} - \frac{\mathrm{i}\sin\phi}{\rho}\psi_\rho\right) \quad (34)$$
$$\chi_y = \mathrm{e}^{\mathrm{i}\phi}\left(\sin\phi\,\psi_{\rho\rho} + \frac{\mathrm{i}\cos\phi}{\rho}\psi_\rho\right)$$

and

$$\Phi_z = \psi_{zz} + \mathrm{i}k\psi_z$$
$$\Phi_x + \mathrm{i}\Phi_y = \mathrm{e}^{\mathrm{i}\phi}\Phi_\rho = \mathrm{e}^{\mathrm{i}\phi}(\psi_{\rho z} + \mathrm{i}k\psi_\rho) \quad (35)$$
$$\chi_x - \mathrm{i}\chi_y = \psi_{\rho\rho} + \psi_\rho/\rho$$
$$\mathrm{i}(\chi_x\chi_y^* - \chi_y\chi_x^*) = (\psi_{\rho\rho}\psi_\rho^* + \psi_\rho\psi_{\rho\rho}^*)/\rho.$$

We shall show that beams with vector potential given by (32) with cylindrically symmetric $\psi$ are exactly circularly polarized on the beam axis $\rho = 0$. These beams are approximately circularly polarized near the axis: we shall see that $\Lambda = \mathrm{O}(\rho^2)$. From (33)–(35) we have

$$\left(\frac{2k^2}{E_0}\right)^2 \boldsymbol{B}^2 = \chi_x^2 + \chi_y^2 + 2(\chi_x + \mathrm{i}\chi_y)\Phi_z - (\Phi_x + \mathrm{i}\Phi_y)^2$$
$$= \mathrm{e}^{2\mathrm{i}\phi}\left\{\psi_{\rho\rho}^2 - \frac{1}{\rho^2}\psi_\rho^2 + 2\left(\psi_{\rho\rho} - \frac{1}{\rho}\psi_\rho\right)(\psi_{zz} + \mathrm{i}k\psi_z)\right.$$
$$\left. - (\psi_{\rho z} + \mathrm{i}k\psi_\rho)^2\right\} \quad (36)$$

$$\left(\frac{2k^2}{E_0}\right)^2 \boldsymbol{B}\cdot\boldsymbol{B}^* = |\chi_x|^2 + |\chi_y|^2 + 2|\Phi_z|^2 + |\Phi_\rho|^2$$
$$+ \Phi_z(\chi_x^* + \mathrm{i}\chi_y^*) + \Phi_z^*(\chi_x - \mathrm{i}\chi_y)$$
$$= |\psi_{\rho\rho}|^2 + \frac{1}{\rho^2}|\psi_\rho|^2 + 2|\psi_{zz} + \mathrm{i}k\psi_z|^2 + |\psi_{\rho z} + \mathrm{i}k\psi_\rho|^2$$
$$+ 2\,\mathrm{Re}\left\{\left(\psi_{\rho\rho}^* + \frac{1}{\rho}\psi_\rho^*\right)(\psi_{zz} + \mathrm{i}k\psi_z)\right\}. \quad (37)$$

Since $\boldsymbol{E} = \mathrm{i}\boldsymbol{B}$ the polarization measure $\Lambda$ is the same for $\boldsymbol{E}$ and $\boldsymbol{B}$; it is

$$\Lambda = \left(\left|\psi_{\rho\rho}^2 - \frac{1}{\rho^2}\psi_\rho^2 + 2\left(\psi_{\rho\rho} - \frac{1}{\rho}\psi_\rho\right)(\psi_{zz} + \mathrm{i}k\psi_z)\right.\right.$$
$$\left.\left. - (\psi_{\rho z} + \mathrm{i}k\psi_\rho)^2\right|\right)\left(2|\psi_{zz} + \mathrm{i}k\psi_z|^2 + |\psi_{\rho z} + \mathrm{i}k\psi_\rho|^2\right.$$

$$+ 2\operatorname{Re}\left\{(\psi_{zz}+\mathrm{i}k\psi_z)\left(\psi_{\rho\rho}^*+\frac{1}{\rho}\psi_\rho^*\right)\right\}$$
$$+ |\psi_{\rho\rho}|^2 + \frac{1}{\rho^2}|\psi_\rho|^2\Bigg)^{-1}. \tag{38}$$

Consider now the behaviour near the beam axis. When $\psi$ is even in $\rho$, as is the case in all the examples in (25) and (26), we can write

$$\psi(\rho, z) = \psi_0(z) + \rho^2\psi_2(z) + \rho^4\psi_4(z) + \mathrm{O}(\rho^6). \tag{39}$$

(For $\psi_{00}$, $\psi_0(z) = \sin(Z)/Z$ where $Z = k(z-\mathrm{i}b)$.) Then (38) gives the indicated proportionality to $\rho^2$ for small $\rho$ (primes indicate differentiation with respect to $z$):

$$\Lambda = \rho^2\frac{|8(\psi_0''+\mathrm{i}k\psi_0'+2\psi_2)\psi_4 - 2(\psi_2'+\mathrm{i}k\psi_2)^2|}{|\psi_0''+\mathrm{i}k\psi_0+2\psi_2|^2} + \mathrm{O}(\rho^4). \tag{40}$$

When dealing with exact solutions of the Helmholtz equation, we also have $(\nabla^2+k^2)\psi = 0$ satisfied to each order in $\rho^2$, so that

$$\psi_2 = -\tfrac{1}{4}(\psi_0''+k^2\psi_0), \qquad \psi_4 = -\tfrac{1}{16}(\psi_2''+k^2\psi_2). \tag{41}$$

Then (40) may be rewritten as

$$\Lambda = \tfrac{1}{4}\rho^2\frac{|2(F''+k^2F)^2 - (F'+\mathrm{i}kF)[(F'-\mathrm{i}kF)''+k^2(F'-\mathrm{i}kF)]|}{|F'+\mathrm{i}kF|^2}$$
$$+ \mathrm{O}(\rho^4) \tag{42}$$

where $F = \psi_0' + \mathrm{i}k\psi_0$.

By way of example, the wavefunction $\psi_{00} = \sin kR/kR$ with $R^2 = \rho^2 + (z-\mathrm{i}b)^2$ has $\psi_0 = \frac{\sin k(z-\mathrm{i}b)}{k(z-\mathrm{i}b)}$. This gives, in the focal plane $z = 0$,

$$\Lambda_0 \to \begin{cases} \tfrac{1}{4}(\rho/b)^2\beta^{-2} + \mathrm{O}(\rho^4) & (\beta \gg 1) \\ \tfrac{1}{40}(k\rho)^2 + \mathrm{O}(\rho^4) & (\beta \ll 1). \end{cases} \tag{43}$$

The complete behaviour in the focal plane is shown in figures 4–6, and limiting analytic expressions are given in appendix B. We see that our steady beam constructed to have circular polarization in the plane wave limit, with vector potential given in (32), is exactly circularly polarized on the beam axis, but that away from the axis the polarization deviates from exactly circular, in proportion to $\rho^2/k^2b^4$ for wide beams, and in proportion to $(k\rho)^2$ for very tightly focused beams ($\beta = kb$ large and small compared to unity, respectively).

## 5. Discussion

We have seen that the finiteness in the transverse directions of actual electromagnetic beams imposes constraints which make impossible the usual properties assumed for plane electromagnetic waves. Substantial deviations from the ideal, almost-plane-wave polarization properties [18] are to be expected in tightly focused beams. In particular, there is a strong constraint on circular polarization: an example is given of a class of beams within which there can be exact circular polarization only on the beam axis, with the deviation from circular polarization being initially in proportion to the square of the distance from the beam axis. The broader the beam, the larger the axial region where the polarization is nearly circular.

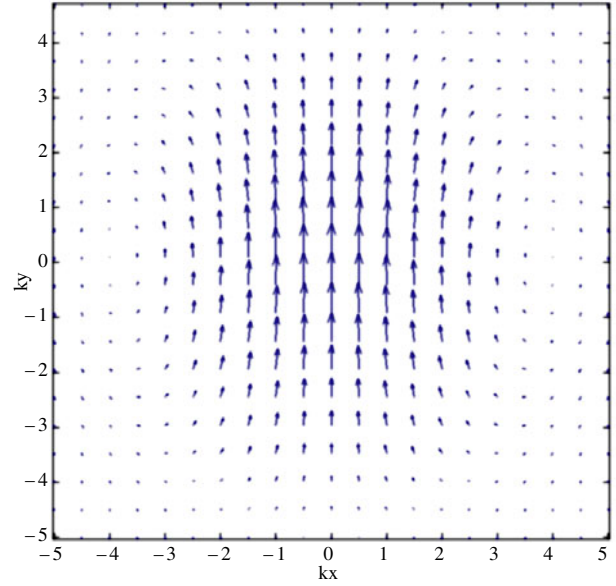

**Figure 4.** Electric field in the focal plane at $t = 0$ for the 'CP' beam with vector potential given by (32) and $\psi = \psi_{00}$ with $\beta = 2$. Over the centre of the figure the field is circularly polarized and so the field vectors rotate with time. The polarization tends to linear in the outer region: see figures 5 and 6.

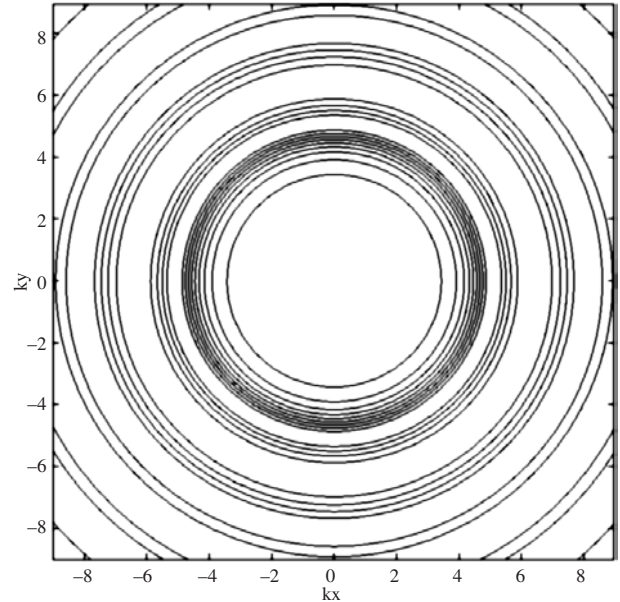

**Figure 5.** Contour plot of the polarization measure $\Lambda$ for the 'CP' beam, with $\psi = \psi_{00}$ and $\beta = 2$. The polarization is exactly circular at the centre of the focal plane, where $\Lambda = 0$. The contours shown are in 10% increments and decrements, from 0.1 to 0.9. There are rings of linear polarization, at $k\rho \approx 5.1, 8.1, \ldots$. The radial dependence of $\Lambda$ is also shown in figure 6.

A degree of linear polarization was introduced in section 3, namely

$$\Lambda = \frac{E_1^2 - E_2^2}{E_1^2 + E_2^2} = \frac{|\boldsymbol{E}^2(\boldsymbol{r})|}{|\boldsymbol{E}(\boldsymbol{r})|^2} \tag{44}$$

where $E_1$ and $E_2$, given by (5), are the lengths of the semiaxes of the polarization ellipse. This polarization measure has the advantage of being simply expressed as the ratio of the magnitude of the square of the complex field to the square of the magnitude. Hurwitz [19] has used the polarization parameter

$$S = \frac{2E_1 E_2}{E_1^2 + E_2^2}. \tag{45}$$

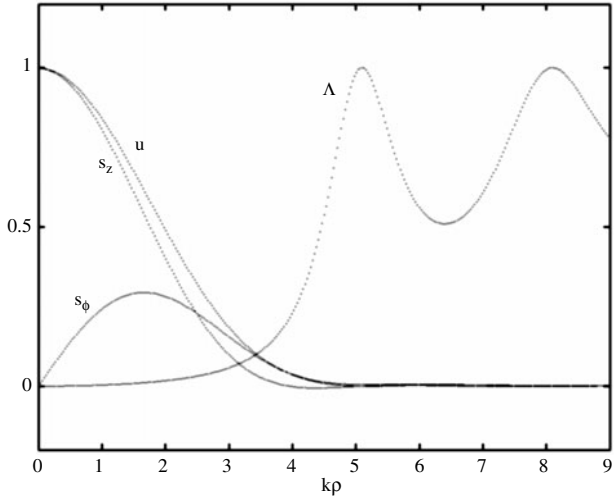

**Figure 6.** Focal plane properties of the 'CP' beam, with $\psi = \psi_{00}$ and $\beta = 2$. The curves show radial dependence of the polarization measure $\Lambda$, the energy density $u$, and the longitudinal ($S_z$) and azimuthal ($S_\phi$) components of the energy flux (Poynting vector) $\mathbf{S} = \frac{c}{4\pi}\mathbf{E} \times \mathbf{B}$. The energy density has been divided by $u_0$, the energy density at the centre of the focal plane. The longitudinal component $S_z$ and the azimuthal component $S_\phi$ have been divided by $cu_0$. Note that the energy velocity has magnitude $c$ at the origin (and everywhere along the beam axis). This follows from the conditions (7) for circular polarization and the fact that the 'CP' beam is *steady* (see equations (37) of [4]).

$S$ is zero for linear polarization and unity for circular polarization, while $\Lambda$ is unity and zero respectively. The two measures are related by $S^2 + \Lambda^2 = 1$. The Hurwitz measure has the advantage that for monochromatic unpolarized light it is uniformly distributed between zero and one. The ratio of the minor axis to the major axis of the polarization ellipse is

$$\frac{E_2}{E_1} = \sqrt{\frac{1-\Lambda}{1+\Lambda}} = \frac{1-\sqrt{1-S^2}}{S} \tag{46}$$

and when $S = \frac{1}{2}$ this takes the value $2 - \sqrt{3} \approx 0.268$. The corresponding $\Lambda$ value is $\sqrt{3}/2 \approx 0.866$. Barakat [20] has extended the Hurwitz results to partially polarized light, and corrected an error in [19].

The polarization measures $\Lambda(\mathbf{r})$ or $S(\mathbf{r})$ characterize the polarization of a monochromatic beam at the point $\mathbf{r}$. They both range from 0 to 1, with the extremes corresponding to circular and linear polarization. Both $\Lambda$ and $S$ can be written in terms of the lengths of the real orthogonal vectors $\mathbf{E}_1(\mathbf{r})$ and $\mathbf{E}_2(\mathbf{r})$ defined in terms of $\mathbf{E}_r(\mathbf{r})$ and $\mathbf{E}_i(\mathbf{r})$ in section 1: see (44) and (45). The plane of the polarization ellipse is the plane of $\mathbf{E}_r, \mathbf{E}_i$ or $\mathbf{E}_1, \mathbf{E}_2$. An equivalent characterization can be given in terms of the Stokes parameters (see for example [2, sections 1.4.2 and 10.8.3], where the Stokes parameters are defined for harmonic plane waves and for quasi-monochromatic plane waves, respectively), provided we understand their definition to be local in the case of beams, as opposed to global (independent of $\mathbf{r}$) in the case of plane waves. For the coherent monochromatic beams considered in this paper, the Stokes parameters (all functions of $\mathbf{r}$) become

$$s_0 = E_r^2 + E_i^2, \qquad s_1 = E_r^2 - E_i^2,$$
$$s_2 = 2\mathbf{E}_r \cdot \mathbf{E}_i, \qquad s_3 = 2\sqrt{E_r^2 E_i^2 - (\mathbf{E}_r \cdot \mathbf{E}_i)^2}. \tag{47}$$

Thus $\Lambda(\mathbf{r})$ and $S(\mathbf{r})$ are given in terms of the Stokes parameters by

$$\Lambda = \sqrt{1 - s_3^2/s_0^2}, \qquad S = s_3/s_0. \tag{48}$$

Barakat [21] and Eliyahu [22] have studied the statistics of the Stokes parameters for partially polarized light; Brosseau [23] relates the Stokes parameters to the coherency matrix formalism. Carozzi *et al* [24] have developed a more general characterization of polarization in terms of the spectral density tensor, and Berry and Dennis [25] have studied polarization singularities (e.g. C lines, where the polarization is purely circular) in isotropic random vector waves. The papers [19–25] are all concerned with the *statistics* of wave fields; that issue has not been treated in the present paper, which is restricted to idealized laser beams: monochromatic and coherent.

## Acknowledgments

The author is indebted to Damien Martin and Tim Benseman for helpful comments. This work has also benefited from the suggestions and references kindly provided by two anonymous referees.

## Appendix A. The polarization of an 'LP' beam, with $\psi = \psi_{00}$

The polarization measure $\Lambda$ for the $\mathbf{A} = A_0(\psi, 0, 0)$ 'linearly polarized' beam is given in (27). The simplest exact beam wavefunction, $\psi_{00} = \sin kR/kR$, $R^2 = \rho^2 + (z - ib)^2$, is best handled in oblate spheroidal coordinates $(\xi, \eta)$, since these are proportional to the real and imaginary parts of $R$ [3, 17]. We have, with $\beta = kb$,

$$\rho = b\sqrt{(1+\xi^2)(1-\eta^2)}, \qquad z = b\xi\eta,$$
$$R = b(\xi - i\eta), \qquad \psi_{00} = \frac{\sin\beta(\xi - i\eta)}{\beta(\xi - i\eta)}. \tag{A.1}$$

The differentiations with respect to $\rho$ and $z$ required in the evaluation of $\Lambda$ can be converted to $\xi$ and $\eta$ differentiations as given in equations (A.1)–(A.5) of [3]. The complete expression for $\Lambda$ (which depends on $\beta$, $\xi$, $\eta$, $\cos\beta\xi$, $\sin\beta\xi$, $\cosh\beta\eta$ and $\sinh\beta\eta$) is too lengthy to be given here. We shall give only the expressions for the polarization evaluated in the focal plane $z = 0$. There are two regions to be considered: the disc $\rho \leqslant b$, and the remainder of the $z = 0$ plane, $\rho \geqslant b$.

These correspond to

$$\{z=0, 0 \leqslant \rho \leqslant b\}: \quad \xi = 0, \quad \eta = \sqrt{1-\rho^2/b^2},$$
$$0 \leqslant \eta \leqslant 1, \quad \rho = b\sqrt{1-\eta^2}$$
$$\{z=0, 0 \geqslant b\}: \quad \eta = 0, \quad \xi = \sqrt{\rho^2/b^2 - 1},$$
$$0 \leqslant \xi < \infty, \quad \rho = b\sqrt{1+\xi^2}. \tag{A.2}$$

Both $\xi$ and $\eta$ are zero on the circle $\{z=0, \rho=b\}$. On this circle $\Lambda$ takes the value

$$\Lambda_c = \frac{10(5+\beta^2 \cos^2\phi)}{50 + \beta^2(\beta^2+10)\cos^2\phi}. \tag{A.3}$$

As expected, this is unity along the $y$-axis, where $\cos\phi = 0$. $\Lambda_c$ is nearly unity for all $\phi$ for small $\beta$, but for large $\beta$ it takes the small value $10/\beta^2$ except in the immediate neighbourhood of $\cos\phi = 0$, where it rapidly increases to unity. The degree of linear polarization within the near region (the disc $\rho \leqslant b$) in the focal plane has the limiting values for large and small $\beta$

$$\Lambda_n = \begin{cases} \dfrac{\eta^2(\cos^2\phi + \eta^2 \sin^2\phi)}{(2-\eta^2)\cos^2\phi + \sin^2\phi} + O(\beta^{-1}) \\ 1 - (1-\eta^2)\beta^4 \cos^2\phi + O(\beta^6). \end{cases} \tag{A.4}$$

Both limits give unity (perfect linear polarization) at the origin, where $\eta \to 1$. In the far region, $\rho \geqslant b$, we find the large and small $\beta$ values to be

$$\Lambda_f = \begin{cases} \dfrac{\xi^4(\cos^2\phi - \xi^2 \sin^2\phi)}{\xi^4 \sin^2\phi + (2+\xi^2)\cos^2\phi} + O(\beta^{-1}) \\ \dfrac{\xi^2[\xi^2 + 3(1+\xi^2)\cos^2\phi]}{\xi^4 + 3(\xi^4 + 7\xi^2 + 6)\cos^2\phi} + O(\beta). \end{cases} \tag{A.5}$$

For any $\beta$, the asymptotic value of $\Lambda$ far from the axis has the leading terms

$$\Lambda_f = 1 - \frac{2\cos^2\phi}{\xi^2 \sin^2\phi} + O(\xi^{-3}). \tag{A.6}$$

Thus for $\rho \gg b$ the beam is nearly linearly polarized in most of the focal plane, except close to the $x$-axis (small $\phi$).

Along the beam axis ($\rho = 0$) the polarization is everywhere exactly linear. Near the axis, the deviation of $\Lambda$ from unity is proportional to $\rho^2 \cos^2\phi$ times a function of $z$.

## Appendix B. The polarization of a 'CP' beam, with $\psi = \psi_{00}$

We proceed as in appendix A, evaluating $\Lambda$ as given in (38) with $\psi = \sin kR/kR$. Again the complete expression for $\Lambda$ is too long to be given here, and we will give only the limiting expressions for the polarization evaluated in the focal plane $z = 0$. In this plane, on the circle $\rho = b$, $\Lambda$ takes the value

$$\Lambda_c = \frac{5\beta^2}{2\beta^4 + 20\beta^3 + 95\beta^2 + 200\beta + 200} = \begin{cases} \dfrac{5}{2\beta^2} + O(\beta^{-3}) \\ \dfrac{\beta^2}{40} + O(\beta^3). \end{cases} \tag{B.1}$$

We see that $\Lambda_c$ is small at large and at small $\beta$; its maximum value of $(27 - 8\sqrt{10})/89 \approx 0.019$ is attained at $\beta = \sqrt{10}$. Thus the beam has at most an ellipticity of $[2\Lambda_c/(1+\Lambda_c)]^{\frac{1}{2}} \approx 0.19$ at $\rho = b$.

The limiting values of $\Lambda$ in the near region within the disc $\rho \leqslant b$ are

$$\Lambda_n = \begin{cases} \dfrac{1-\eta}{1+\eta}\dfrac{1}{\beta^2} + O(\beta^{-1}) \\ \dfrac{1}{40}(1-\eta^2)\beta^2 + O(\beta^3). \end{cases} \tag{B.2}$$

The small-$\beta$ limit is the same as the leading term in (43), since $(1-\eta^2)\beta^2 = (k\rho)^2$. The large-$\beta$ limit also agrees with that in (43), since $\frac{1-\eta}{1+\eta} = \frac{\rho^2}{4b^2} + O(\rho^4)$.

The limiting values of $\Lambda$ in the far region $\rho \geqslant b$ are

$$\Lambda_f = \begin{cases} \dfrac{\xi^2(1+\xi^2)}{\xi^4 + (1+2\cos^2\beta\xi)\xi^2 + (4\cos\beta\xi\sin\beta\xi)\xi + 2\sin^2\beta\xi} \\ \quad + O(\beta^{-1}) \\ \dfrac{1}{40}(1+\xi^2)\beta^2 + O(\beta^3). \end{cases} \tag{B.3}$$

For any $\beta$, the asymptotic value of $\Lambda$ far from the axis is unity:

$$\Lambda_f = 1 - \frac{2(\beta+2)^2 \cos^2\beta\xi}{\beta^2 \xi^2} + O(\xi^{-3}). \tag{B.4}$$

Thus the 'circularly polarized' beam tends to linear polarization far from the axis (for $\rho \gg b$)! Away from the focal region, however, the polarization is largely circular: the linear polarization in the far field is restricted to the region $\cos\theta \ll \beta^{-1}$, where $\theta$ is the angle to the beam axis.

The energy density for a steady beam can be written as $u = \boldsymbol{B} \cdot \boldsymbol{B}^*/8\pi$ (see [4, equation (58)]), and for the 'CP' beam the required scalar product is given in (37). When $\psi = \psi_{00}$ we find that, in the focal plane, $u \sim \rho^{-2}$ for $\rho \gg b$. Thus the integral of the energy density over the focal plane diverges logarithmically, as was shown to be the case for the TM, TE and 'TEM' beams with $\psi = \psi_{00}$ [3].

The energy flux (Poynting vector) for a steady beam can be written as $\boldsymbol{S} = (c/8\pi)i\boldsymbol{B} \times \boldsymbol{B}^*$ (see [4, equation (58)]). When $\psi$ is independent of the azimuthal angle $\phi$, the longitudinal component $S_z$ is independent of $\phi$, and the transverse components resolve into radial and azimuthal parts $S_\rho$ and $S_\phi$, both of which are independent of $\phi$: $S_x = S_\rho \cos\phi - S_\phi \sin\phi$, $S_y = S_\rho \sin\phi + S_\phi \cos\phi$ (as is the case for the 'TEM' beam: [4, see section 6]). In the focal plane, and with $\psi = \psi_{00}$, $S_\rho$ is zero, and $S_z$ and $S_\phi$ have the asymptotic forms

$$S_z \sim \frac{\cos k\rho \sin k\rho}{\rho^3} + O(\rho^{-4}),$$
$$S_\phi \sim \frac{\cos^2 k\rho}{\rho^3} + O(\rho^{-4}). \tag{B.5}$$

Thus the momentum integral (proportional to $\int_0^\infty d\rho\, \rho S_z$) is convergent, as it is for the TM, TE and 'TEM' cases [3, 26]. The angular momentum density is $c^{-2}\rho S_\phi$ (see [4, equation (64)]), and thus for $\psi = \psi_{00}$ the total angular momentum for the 'CP' beam is logarithmically divergent. These results reinforce the statement that 'the usefulness of $\psi_{00}$ appears to be limited to the region near the beam axis' [26].